\documentclass[aps,10pt,onecolumn, amsmath,amssymb,groupedaddress, showkeys]{revtex4-2}
\usepackage{graphicx}
\usepackage{dcolumn}
\usepackage{bm}
\usepackage[normalem]{ulem} 
\usepackage{color}
\usepackage{float}
\usepackage{caption}
\usepackage{subcaption}
\DeclareCaptionJustification{justified}{\leftskip=0pt \rightskip=0pt \parfillskip=0pt plus 1fil}
\usepackage{lineno}
\usepackage{epstopdf}
\usepackage{stmaryrd}
\usepackage{upgreek}
\begin{document}
	\title{Linear poroelastic response of thin permeable gel films}
	\author{Caroline Kopecz-Muller$^{1,2,3,4,\bm{*}}$}
	\author{Joshua D. McGraw$^{2,3,\bm{*}}$}
	\author{Thomas Salez$^{4,\bm{*}}$}
	\affiliation{$^1$Laboratoire Navier (Ecole des Ponts Paris Tech - Université Gustave Eiffel - CNRS), 77420 Champs-sur-Marne, France.}
	\affiliation{$^2$Gulliver, CNRS UMR 7083, ESPCI Paris, Université PSL, 75005 Paris,  France.}
	\affiliation{$^3$Institut Pierre-Gilles de Gennes, 75005 Paris, France.}
	\affiliation{$^4$Univ. Bordeaux, CNRS, LOMA, UMR 5798, 33405 Talence, France.}
	\date{\today}
	\begin{abstract}
		When a hydrophilic and deformable porous material is immersed in a bath, it may absorb the solvent and expand by several times its volume, thus forming a highly soft and porous hydrogel. A stress applied on the soft hydrogel surface deforms it and forces the absorbed solvent to move by flowing through the network of pores. This coupled phenomenon sets the framework of poroelasticity. Moreover, polymeric gels are often used in ultra-thin coatings to tune surface properties. Together with the characteristic poroelastic coupling, this thinness challenges the modelling of their response. In this article, we derive the point-force mechanical response of a thin, permeable and poroelastic layer bounded to a rigid substrate. We show that the gel surface is only deformed around the indentation point, within a radius on the order of the layer thickness. The obtained Green's function can be directly used to predict the space- and time-dependent surface deformation of the gel. Our findings are relevant for a broad range of applications, such as indentation experiments on swollen gels, thin membranes or soft and living systems, as well as lubrication problems involving a soft and porous wall, for instance in microfluidics.
	\end{abstract}
	\maketitle
	
	\small\textbf{$\bm{^*}$Corresponding authors:} caroline.kopecz-muller@univ-eiffel.fr, joshua.mcgraw@cnrs.fr, thomas.salez@cnrs.fr \normalsize
	\section{introduction}

Soft and permeable matter is ubiquitous in living systems: emblematic examples are found among sea sponges~\cite{reiswig1974water}, bacteria organelles~\cite{greening2020formation}, cartilaginous joints~\cite{hou1992analysis, jahn2016lubrication} and eyelid wipers~\cite{cher2008new}. Such soft permeability can also be observed in man-made objects, for instance in foamy materials~\cite{cohen2013flow, yang2019multiaxial} and grafted polymer layers~\cite{russell2002surface}. The latter example has been extensively used in coatings for tuning surface properties, promoting cellular culture~\cite{bearinger1997p} or enhancing lubrication~\cite{davies2018elastohydrodynamic,bureau2023lift}, but also building microfluidic valves~\cite{beebe2000functional, idota2005microfluidic}, single-cell traps~\cite{d2018microfluidic}, or cages for drug delivery~\cite{peppas2006hydrogels}. Such engineering efforts would benefit from an \emph{a-priori} understanding of the mechanical response of the considered material that constitutes the coating. For purely elastic matter, this question has been addressed in various geometries, including the axial loading on a flat layer~\cite{li1997elastic}, or the contact between two elastic spheres~\cite{hughes1979soft}. Then, the resulting purely elastic response has been used to describe droplet settling, extensively for soft-lubrication problems~\cite{essink2021regimes, leroy2011hydrodynamic,rallabandi2024fluid,wang2017elastic,farzaneh2014hydrodynamic}, and the associated lift force in elastohydrodynamic (EHD) situations~\cite{salez2015elastohydrodynamics, skotheim2005soft, bertin2022soft, pandey2016lubrication, bureau2023lift}. To aim at a more accurate description of real systems, efforts have been made to take into account additional features, such as surface tension~\cite{hu2023effect}, and viscoelasticity~\cite{hu2023effect, kargar2021lift, rallabandi2024fluid} of the solid boundary.   
	
Materials that are used in the aforementioned soft-lubrication contexts are often polymeric~\cite{bureau2023lift, davies2018elastohydrodynamic, wang2017elastic,wang2017elastic2,wang2015out}. They may thus become porous when reticulated to form a network and exposed to a solvent. This solvent may penetrate into the polymeric matrix through the pores, depending on the solvent-polymer affinity~\cite{flory1953principles,rubinstein2003polymer, doi2013soft}. While polymer solutions are known to slip at walls~\cite{leger1997wall,mhetar1998slip,huang2008water,bocquet2010nanofluidics}, a similar feature is demonstrated for walls made of reticulated polymer gels. The gel porosity induces a slip length that can be estimated to be on the order of the pore size~\cite{beavers1967boundary}. A first approach to model the lubrication problem near such a porous wall is to make the assumption that the gel does not deform, taking into account only its porosity~\cite{knox2017squeeze}. However, polymer gels are often soft materials, with Young Modulii in the kPa to MPa range. A coupling thus arises between elasticity and fluid flow through the porous matrix. 
	
The question of poroelasticity was first addressed by Biot in the 1940's, in the context of soil sedimentation~\cite{biot1941general,biot1956general}. In Biot's work, soil elasticity is modelled by a classical stress-strain relationship given by a generalized Hooke's law, and an assumption of small deformations ensures the linearity of the problem~\cite{love2013treatise,landau2012theory}. Later works complemented that of Biot, with focuses on different geometries~\cite{rice1976some, wang2000theory}. On the fundamental side, a recent experimental interest for gels~\cite{jeon2023moving, etzold2021transpiration, delavoipiere2018friction, ciapa2020transient, delavoipiere2016poroelastic, cuccia2020pore, hu2010using, zhao2018geometrical, kopecz2025swelling}, or fiber poroelasticity~\cite{van2022spontaneous,van2025spreading}, has emerged, following Biot's linear approach. For the ideal isotropic case of gel spheres swelling in a solvent, another approach consists of modelling the elasticity of the polymeric network by adding an elastic contribution to the classical free energy density from the Flory-Huggins polymer theory~\cite{flory1953principles, onuki2005theory}. This method leads to the development of a so-called Terzaghi stress~\cite{terzaghi1936shearing}. In this framework, the effective elastic modulus and porosity (related to the volume fraction of voids within the polymeric matrix) of the gel depend on the solvent fraction~\cite{sierra2011bulk, tokita1991friction,engelsberg2013free}. This strongly non-linear approach is suitable to describe problems with spherical symmetry, in large deformations~\cite{bertrand2016dynamics, macminn2016large, style5092714characterizing}. A recent and promising alternative strategy to Terzaghi stress considered deviatoric strain to model large-deformation swelling~\cite{webber2023linear1, webber2023linear2}. Another approach is based on Biot's method and investigates the competition between the dissipation due to the fluid viscosity and the friction within the pores~\cite{hu2012viscoelasticity}. In summary, if an evolution equation on the solvent fraction is derived in each particular case to describe the swelling dynamics, isolating the pure poroelastic response of the gel in a general case, namely the Green's function, remains a challenge to address. 
	
In a previous work, we described the linear mechanical response of a permeable poroelastic half-space subject to a point-force pressure source and extended it to any axisymmetric pressure field. In the framework of linear poroelasticity, we combined the classical continuum mechanics description of an elastic solid to Darcy's law, expressing fluid flows inside a porous medium, driven by chemical-potential gradients. The calculation is detailed in~\cite{kopecz2023mechanical} and~\cite{kopecz2024mechanics} and leads to the Green's function of the problem, which mathematically expresses the deformation of the surface in space and time. However, the latter solution only reflects ideal situations of thick poroelastic objects. %, and raises the question of how thick should be the material to be considered as semi-infinite. 
In the present work, we follow the same calculation path as before, but taking into account finite-thickness of the porous layer. We derive the corresponding poroelastic Green's function for thin, permeable layers. For this finite case, we show that the mechanical response of the layer extends around the application point, with a region with radius of order the film thickness, but vanishes in the extended region beyond. 
	
The present article is organized as follows: in Sec.~\ref{sec:poroelasticity}, we recall the mathematical framework of linear poroelasticity (Sec.~\ref{sec:eq_poroelasticity}) and translate the finite-thickness feature into equations (Sec.~\ref{sec:BC}). In Sec.~\ref{sec:resolution}, we briefly exhibit the resolution technique that is used, similarly to~\cite{kopecz2023mechanical}. Results are exhibited in Sec.~\ref{sec:results}: we first derive the deformation profile in time of a permeable poroelastic thin layer in reciprocal space (Sec.~\ref{sec:green}). Then, we study the asymptotics to physically interpret the behaviour or the material (Sec.~\ref{sec:asymptotics}) and demonstrate the continuity with our previous work on semi-infinite poroelastic media (Sec.~\ref{sec:thin_thick}). To validate our interpretations, we compute the deformation back to real space (Sec.~\ref{sec:real}). Finally, we extend our results to arbitrary pressure fields having the same symmetry properties (Sec.~\ref{sec:extention}), and we conclude.

	\section{Linear poroelasticity framework}
	\label{sec:poroelasticity}
	\subsection{Poroelasticity equations}
	\label{sec:eq_poroelasticity}
\begin{figure}
		\centering
		\includegraphics[width= 0.5 \linewidth]{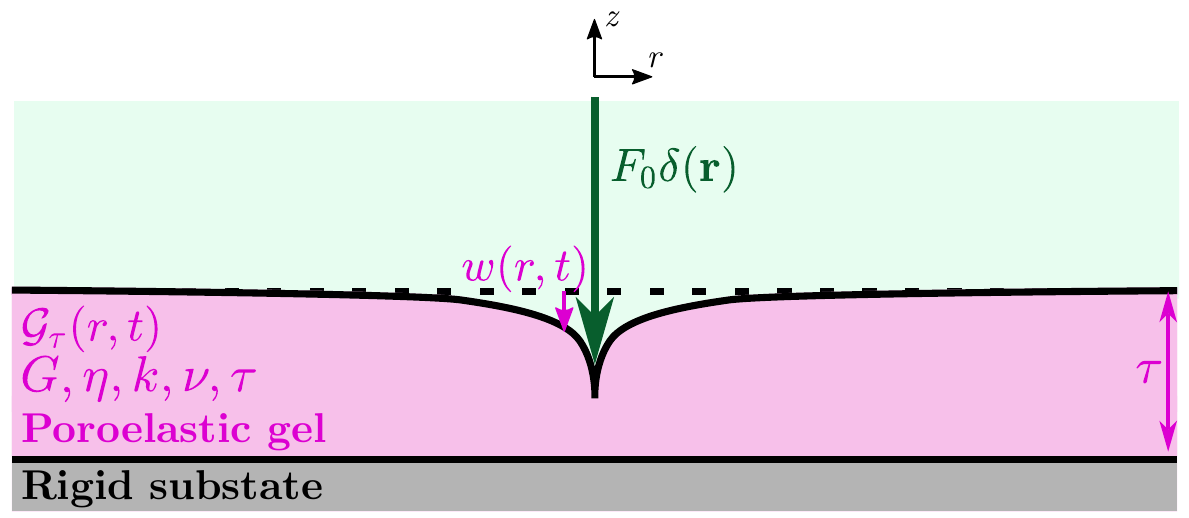}
		
		\caption{\textbf{Schematic of the system.} \textit{A point force $F_0\delta(\bm{r})$ is applied at the surface of a permeable poroelastic medium, of shear modulus $G$ and Poisson ratio $\nu$. A fluid of viscosity $\eta$ is flowing through the pores of the gel, whose permeability is denoted by $k$. The resulting vertical deformation profile represented by the thick black line is denoted by $w(r,t)$. We compute the poroelastic response function $\mathcal{G}_\tau(r,t)$ of the material.}}
		\label{fig:schematic}
	\end{figure}

	The physical situation is represented in Fig.~\ref{fig:schematic}. We model a gel bound to a substrate and immersed in its own solvent by a poroelastic layer located between $z = -\tau$ and $z=0$, with the vertical coordinate $z$ and $\tau$ the thickness of the layer. We suppose that the mechanical response of the gel is described by linear poroelasticity equations, that we introduce in the following.
	We use the classical framework of continuum mechanics, restraining ourself to small deformations. Thus, the strain tensor $\bm{\epsilon}$ is linearly expressed from the symmetric part of the displacement field gradient tensor, as :
	\begin{equation}
		\bm{\epsilon} = \frac{1}{2}\left[ \bm{\nabla}\mathbf{u} + (\bm{\nabla}\mathbf{u})^T\right],
		\label{eq:strain-def}
	\end{equation}
	with $\mathbf{u}$ referring to the displacement field with respect to the reference state. In the framework of linear poroelasticity, a term is added to the classical expression that links the stress tensor $\bm{\sigma}$ to the strain tensor $\bm{\epsilon}$, namely Hooke's law in a linear and isotropic case. Often referred to as the pore pressure, this extra term invokes the variations of solvent chemical potential $\mu$ with respect to the equilibrium value $\mu_0$, and expresses the extra stress locally exerted on the solid body by inhomogeneities of solvent. The stress-strain relationship reads :
	\begin{equation}
		\bm{\sigma}=2G\Big[\bm{\epsilon} +\frac{\nu}{1-2\nu}\text{Tr}(\bm{\epsilon})\mathbf{I}\Big]-\frac{\mu-\mu_0}{\Omega}\mathbf{I},
		\label{eq:stress-strain}
	\end{equation}
	where $\mathbf{I}$ and $\text{Tr}$ are the identity tensor and the trace operator, respectively, while $G$, $\nu$ and $\Omega$ stand for the shear modulus, Poisson ratio and molecular volume of solvent. Fluid flows through the solid structure are described by Darcy's law, stating that the solvent flux  $\mathbf{J}$ is driven by the gradient of chemical potential, as :
	\begin{equation}
		\mathbf{J} = -\Big(\frac{k}{\eta \Omega^2} \Big) \bm{\nabla} \mu,
		\label{eq:darcy}
	\end{equation}
	where $\eta$ and $k$ stand for the solvent viscosity and the permeability of the gel, respectively. The solvent flow follows the mass conservation equation, as:
	\begin{equation}
		\displaystyle \frac{\partial c}{\partial t} + \bm{\nabla}\cdot \mathbf{J}=0,
		\label{eq:adv}
	\end{equation}
	with $c$ the solvent concentration. The latter is considered incompressible. However, concentration variations lead to volume variation of the whole gel, which is effectively compressible. The solvent incompressibility condition reads:
	\begin{equation}
		\text{Tr}(\bm{\epsilon})=\bm{\nabla}\cdot \mathbf{u}=(c-c_0)\Omega,
		\label{eq:div-def}
	\end{equation}
	where $c_0$ refers to the equilibrium solvent concentration. Finally, all body forces are balanced at mechanical equilibrium. This is expressed by Navier's closure equation, as: 
	\begin{equation}
		\bm{\nabla}\cdot\bm{\sigma}=\mathbf{0}.
		\label{eq:navier}
	\end{equation}
	
	\subsection{Point force application on a grafted gel}
	\label{sec:BC}
	In this subsection, we express the initial and boundary conditions describing the sudden application of a point force on a finite-size grafted poroelastic medium. We consider the gel at rest for negative times. At $t=0$, by continuity, one has:
	\begin{equation}
		c(r,z,t=0) = c_0.
	\end{equation} 
	 At time $t=0^+$, a punctual force of intensity $F_0$ is suddenly applied normally to the free surface, as:
	\begin{equation}
		\bm{\sigma} \cdot \mathbf{e}_z = -F_0 \delta(\mathbf{r})H(t)  \mathbf{e}_z,
		\label{eq:DELTA}
	\end{equation}
	where $H(t)$ denotes the Heaviside step function and $\delta(\bm{r})$ the Dirac distribution. Shear stress is neglected. We allow fluid exchanges between the reservoir of solvent located at $z>0$, whose chemical potential is set at the equilibrium value $\mu_0$, and the gel through the free surface. This permeability condition is expressed by:
	\begin{equation}
		\mu(r,z=0,t) = \mu_0. \\
		\label{eq:BC_chempot}
	\end{equation}
	
	At the interface with the substrate, located at $z = -\tau$, the gel is bounded to a rigid boundary. Thus, the displacement field is set to zero, as:
	\begin{equation}
		\mathbf{u}(z=-\tau) = \mathbf{0}.
		\label{eq:BC_strain_field}
	\end{equation}
	Yet, the stress field does not necessarily vanish at $z=-\tau$. The rigid boundary is impermeable, thus no exchange of solvent is allowed at the interface between the gel and the rigid substrate. The resulting impermeable boundary condition at the bottom of the gel reads~\cite{zhao2018geometrical}:
	\begin{equation}
		\frac{\partial\mu}{\partial z}(z = -\tau) = 0.
		\label{eq:BC_chempot_finite}
	\end{equation}

	\section{Resolution in spectral domain}
	\label{sec:resolution}
	We use the same resolution technique as in~\cite{kopecz2023mechanical}. We consider the problem in the spectral domain. Specifically, we use the Hankel transform of $j$-th order in space and the Laplace transform in time, with $j\in \{0,1\}$. In such a framework, a given field $X(r,t)$ is transformed into:
	\begin{equation}
		\hat{X}(s,q) =\int_{0}^{\infty} \text{d}t \, e^{-q t}\int_0^\infty \text{d}r\text{ } X(r,t)  r J_j(sr),
		\label{eq:space_forward}
	\end{equation}
	where $J_j$ is the Bessel function of the first kind and $j$-th order. The inversion formula reads:
	\begin{equation}
		\displaystyle X(r,t) = \frac{1}{2\pi i}  \int_{\gamma -i \infty}^{\gamma +i\infty} \text{d}q \, e^{q t} \int_0^\infty \text{d}s\, \hat{X}(s,q)  s J_j(sr),
		\label{eq:inversion}
	\end{equation}
	where the inverse Laplace transform is written using the Bromwich integral. Concerning the order of the Hankel transforms, we note that the shear component $\sigma_{rz}$ of the stress tensor and the radial component $u_r$ of the displacement field are transformed using order $j=1$. In contrast, the two normal components $\sigma_{rr}$ and $\sigma_{zz}$ of the stress tensor, the vertical component $u_z$ of the displacement field, the solvent concentration and chemical potential fields $c$ and $\mu$ are transformed using order $j=0$.
	
	Since the derivation here follows closely that of what was already reported in~\cite{mcnamee1960plane, mcnamee1960displacement,kopecz2023mechanical,kopecz2024mechanics}, we show the details of the calculation in the Supplementary Information, Sec.~\ref{sec:SI-detail-calculation}. Briefly, the strategy is to introduce the two displacement potential functions. Using these and combining Eqs.~(1-6) expressed in Sec.~\ref{sec:eq_poroelasticity}, we derive two uncoupled ordinary differential equations of order 2 and 4 on these potential functions in the spectral domain. Solving these two equations leads to expressions for the two potential functions, using six integration constants that are determined using the boundary conditions from Sec.~\ref{sec:BC}. Finally, the displacement field, stress, strain, solvent concentration and chemical potential are derived in reciprocal space from these two displacement potential functions. In the next section, we exhibit the mathematical expression of the deformation profile of a permeable, finite-thickness poroelastic medium.
	
	\section{Results and discussion}
	\label{sec:results}
	The expression of the poroelastic Green's function of a finite-size permeable grafted gel is given in reciprocal space, following the transformation expressed by Eq.~\eqref{eq:space_forward}.
	
	\subsection{Poroelastic Green's function for thin films}
	\label{sec:green}
	For convenience, we first introduce the following dimensionless auxiliary variables: 
	\begin{subequations}
		\label{eq:dimensionless_variable_finite}
		\begin{align}\label{eq:dimensionless_variable_finite_chi}
			\displaystyle \chi & = \frac{\mathcal{D}_\mathrm{pe}s^2}{q},\\
			\label{eq:dimensionless_variable_finite_beta}
			\displaystyle \beta& = \sqrt{1+\frac{q}{\mathcal{D}_\mathrm{pe}s^2}} = \sqrt{1+\frac{1}{\chi}},\\
			\label{eq:dimensionless_variable_finite_thickness}
			\displaystyle \zeta&= s\tau.
		\end{align}
	\end{subequations}
	The variable $\chi$ given by Eq.~\eqref{eq:dimensionless_variable_finite_chi} is interpreted as a self-similar diffusive variable, with $s$ and $q$ being the reciprocal space and time variables in Hankel and Laplace spaces, respectively. The effective poroelastic diffusion coefficient $\mathcal{D}_\mathrm{pe}$ naturally comes out when combining the Eqs.~(1-6), as detailed in the Supplementary Material, Sec.~\ref{sec:SI-detail-calculation}. The variable $\beta$ given by Eq.~\eqref{eq:dimensionless_variable_finite_beta} is introduced for convenience. Lastly, the variable $\zeta$ given by Eq.~\eqref{eq:dimensionless_variable_finite_thickness} rescales the spatial frequency $s$ with the thickness $\tau$.
	
		\begin{figure}
		\centering
		\includegraphics[width=0.95\linewidth]{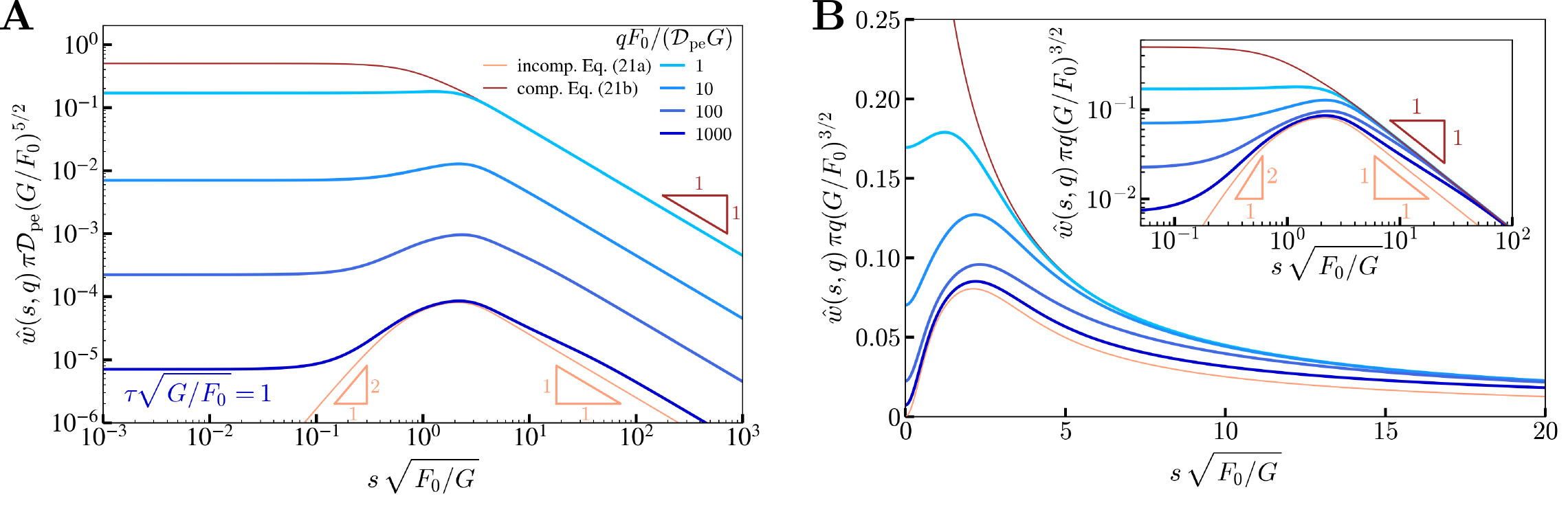}
		
		\caption{\textbf{Deformation profile of a finite-size poroelastic medium in reciprocal space triggered by a point force.}\textit{ \textbf{A}: Deformation $\hat{w}_\tau(s,t)$ scaled by $(F_0/G)^{5/2}/(\pi \mathcal{D}_\text{pe})$, as a function of the dimensionless spatial frequency $s\sqrt{F_0/G}$, for different dimensionless temporal frequencies $qF_0/(\mathcal{D}_\text{pe}G)$, as defined by Eq.~\eqref{eq:green_finite_terms}. \textbf{B}: Deformation multiplied by the temporal frequency $\hat{w}_\tau(s,t)q$, scaled by $ (F_0/G)^{3/2}/\pi$, as a function of the dimensionless spatial frequency $s\sqrt{F_0/G}$, for different dimensionless temporal frequencies $qF_0/(\mathcal{D}_\text{pe}G)$, in linear (main panel) and logarithmic (inset) scales. The known compressible and incompressible limits are represented in brown and orange, as defined in Eqs.~\eqref{asymptotereciprocal_finite}. The dimensionless thickness is set to $\tau\sqrt{G/F_0} = 1$ and the Poisson ratio to $\nu = 0.1$.}}
		\label{fig:green_reciprocal_time}
	\end{figure}
	
	We find the surface normal deformation of the gel $\hat{w}_\tau(s,q)$ by taking the opposite of the vertical displacement, as:
	\begin{equation}
		\displaystyle \hat{w}_\tau(s,q)  = -\hat{u}_z(s,z=0,q) =  \frac{F_0}{4\pi Gsq}\,\frac{N_1+N_2+N_3}{D_1+D_2+D_3},
		\label{eq:green_finite}
	\end{equation}
	with the six dimensionless terms $N_1,N_2,N_3,D_1,D_2,D_3$ expressed as:
	\begin{equation}\left\{
		\begin{array}{l}
			\displaystyle N_1 =  -4\Lambda\beta \chi\sinh(\zeta)\\
			\displaystyle N_2 = \left[(\Lambda \chi +1)\sinh(2\zeta)-2\zeta\right] \beta \cosh(\beta \zeta) \\
			\displaystyle N_3 = \Lambda(3\chi+2 -\chi\cosh(2\zeta))\sinh(\beta \zeta)\\
			\displaystyle D_1 = 4\Lambda \beta \chi \left[\zeta\sinh(\zeta) -(1+\Lambda \chi)\cosh(\zeta) \right]\\
			\displaystyle D_2 = \beta\left[2\zeta^2 + \Big(1+2\Lambda \chi(\Lambda \chi +1)\Big)\Big(1+\cosh(2\zeta)\Big)\right]\cosh(\beta \zeta)\\
			\displaystyle D_3 = -\Lambda(2\chi+1)\left[2\zeta +(\Lambda \chi +1)\sinh(2\zeta)\right]\sinh(\beta \zeta).\\
		\end{array}\right.
		\label{eq:green_finite_terms}
	\end{equation}
	The Poisson ratio $\nu$ appears in the compressibility factor $\Lambda$, defined as:
	\begin{equation}
		\Lambda = \frac{1-2\nu}{1-\nu}. 
		\label{eq:LambdaDef}
	\end{equation}
Finally, the Green's function is deduced in reciprocal space from the deformation scaled by the force amplitude, as : 
	\begin{equation}
	\displaystyle \hat{\mathcal{G}}_\tau(s,q)  = \frac{\hat{w}_\tau(s,q)}{F_0} =  \frac{1}{4\pi Gsq}\,\frac{N_1+N_2+N_3}{D_1+D_2+D_3},
	\label{eq:green_finite_def}
\end{equation}
which expresses the surface deformation in time induced by a point force in reciprocal space.

	To allow a representation using dimensionless variables, length is normalized by $\sqrt{F_0/G}$ and time by $F_0/(G \mathcal{D}_\text{pe})$. After applying Hankel and Laplace transforms that modify the dimension, the deformation $\hat{w}_\tau$ is made dimensionless when scaled by $(F_0/G)^{5/2}/(\pi \mathcal{D}_\text{pe})$ (see SI Sec.~\ref{sec:SI-non-dim}). The former deformation is represented as a function of the spatial frequency $s$ for different temporal frequencies $q$ in Fig.~\ref{fig:green_reciprocal_time}A. For all temporal frequencies $q$, we see first that the deformation is constant at small spatial frequency $s$. This observation applies to the compressible case as well. However, the incompressible case is an exception. Second, we consistently observe a power-law decay with a slope of -1 for large spatial frequencies $s$. To rationalize these observed power-law behaviours, we explore the asymptotics of the deformation expression in the next subsection.
		
	\subsection{Asymptotic analysis}

	\label{sec:asymptotics}	\noindent \textbf{Incompressible and compressible limits :~}
	We first note that if the gel is nearly incompressible, \textit{i.e.} as $\nu \rightarrow 1/2$, and thus $\Lambda \rightarrow 0$, the terms $N_1,N_3,D_1,D_3$ are zero and the surface deformation of the gel reads:
	\begin{equation}
		\label{eq:lim_finite_incompressible}
		\displaystyle \hat{w}_\tau(s,q) \underset{\nu \rightarrow 1/2}{=}  \frac{F_0}{4\pi Gsq}\, \frac{\sinh(2\zeta)-2\zeta}{1+2\zeta^2+\cosh(2\zeta)} =  \frac{1}{q}\hat{w}_\tau^{\mathrm{incomp}}(s),
	\end{equation}
	with $\zeta = s\tau$. Hence, we recover the Laplace transform of the result known for a purely elastic layer of thickness $\tau$ denoted  $\hat{w}_\tau^{\mathrm{incomp}}(s)$~\cite{leroy2011hydrodynamic,li1997elastic}. Similarly, if the permeability is small, \textit{i.e.} as $k\rightarrow 0$, the diffusion constant $\mathcal{D}_\mathrm{pe}$ of the solvent vanishes, $\chi\rightarrow 0$ and $\beta \sim \chi^{-1/2}$, and we recover the same limit. The medium again behaves as an incompressible elastic finite-sized layer. In the opposite limit of large permeability, \textit{i.e.} large $\mathcal{D}_\mathrm{pe}$, $\chi\rightarrow +\infty$ and $\beta\rightarrow 1$, the surface deformation of the gel is finite and reads:
	\begin{equation}
		\label{eq:lim_finite_compressible}
		\displaystyle \hat{w}_\tau(s,q) \underset{k \rightarrow \infty}{=}  \frac{F_0(1-\nu)}{2\pi Gsq}\,\frac{2\zeta -(4\nu-3)\sinh(2\zeta)}{5+4\nu(2\nu-3)+2\zeta^2 -(4\nu-3)\cosh(2\zeta)} = \frac{1}{q}\hat{w}_\tau^{\mathrm{comp}}(s),
	\end{equation}
	where we recover the Laplace transform of the known result for a purely elastic compressible layer of thickness $\tau$, denoted by $ = \hat{w}_\tau^{\mathrm{comp}}(s,q)$, for any Poisson ratio $\nu$~\cite{leroy2011hydrodynamic,li1997elastic}.
	\newline 
	
	\noindent \textbf{Initial and final responses :~}	On Fig.~\ref{fig:green_reciprocal_time}A, we observe that the deformation profiles range from the compressible to the incompressible limits with increasing temporal frequency $q$. To explain this behaviour, we explore the temporal asymptotics of the governing Eqs.~\eqref{eq:green_finite} and~\eqref{eq:green_finite_terms}. We find:
	\begin{subequations}
		\label{asymptotereciprocal_finite}
		\begin{equation}
			\label{eqshort_finite}
			\displaystyle \hat{w}_\tau(s,q) \underset{q \rightarrow \infty}{\sim}  \frac{F_0}{4\pi Gsq}\, \frac{\sinh(2\zeta)-2\zeta}{1+2\zeta^2+\cosh(2\zeta)} = \frac{1}{q}\hat{w}_\tau^{\mathrm{incomp}}(s),
		\end{equation}
		\begin{equation}
			\label{eqlong_finite}
			\displaystyle \hat{w}_\tau(s,q) \underset{q \rightarrow 0}{\sim}  \frac{F_0(1-\nu)}{2\pi Gsq}\,\frac{2\zeta -(4\nu-3)\sinh(2\zeta)}{5+4\nu(2\nu-3)+2\zeta^2 -(4\nu-3)\cosh(2\zeta)} = \frac{1}{q}\hat{w}_\tau^{\mathrm{comp}}(s),
		\end{equation}
	\end{subequations}
	with $\zeta = s\tau$. Equations~\eqref{asymptotereciprocal_finite} are plotted in Fig.~\ref{fig:green_reciprocal_time}(a) and Fig.~\ref{fig:green_reciprocal_time}(b), respectively. We note that we recover the same results computed in the small and large permeability limits respectively, which are the known responses of a purely elastic, finite-thickness layer, respectively incompressible and compressible ~\cite{leroy2011hydrodynamic,li1997elastic}. By invoking the initial and final-value theorems in the short-time and long-time limits of the surface deformation, we find:
	\begin{subequations}
		\label{eq:initial_final_value_thm_finite}
		\begin{align}
			\label{eq:initial_value_thm_finite}
			\displaystyle \hat{w}_\tau(s,t=0^+) &= \lim_{q \rightarrow \infty\,}\, q\hat{w}_\tau(s,q) = \hat{w}_\tau^{\mathrm{incomp}}(s),\\
			\label{eq:final_value_thm_finite}
			\displaystyle \hat{w}_\tau(s,t\rightarrow\infty) &= \lim_{q \rightarrow 0^+} q\hat{w}_\tau(s,q) =  \hat{w}_\tau^{\mathrm{comp}}(s).
		\end{align}
	\end{subequations}
	Thus, at initial times, we recover the response to a point force of a purely elastic incompressible layer, of shear modulus $G$ and thickness $\tau$. At long times, we have the response to a point force of a purely elastic compressible layer, of shear modulus $G$, thickness $\tau$ and Poisson ratio $\nu$.
	
	The incompressible and compressible limits are added in Fig.~\ref{fig:green_reciprocal_time}A. With increasing temporal frequency $q$ (\textit{i.e.} decreasing time), we indeed observe that the deformation profiles range from the compressible thus final asymptote, to the incompressible, thus initial one. Then, given the $1/q$ prefactor appearing in Eqs.~\eqref{eq:green_finite_terms} and~\eqref{asymptotereciprocal_finite} as a natural result of performing a Laplace transform, the dimensionless deformation $\hat{w}_\tau$ multiplied by the temporal frequency $q$ is represented in Fig.~\ref{fig:green_reciprocal_time}B as a function of the spatial frequency $s$. In this representation, in the small-$s$ limit, we observe that the profiles range from compressible to the incompressible limits with increasing spatial frequencies. We then obtain a collapse of the deformation profiles for different temporal frequencies in the limit of large $s$, following the compressible asymptote. We now rationalise this observation by computing the small and large-$s$ limits.  
	\newline
	
	\noindent\textbf{Central and peripheral responses :~}
	We now explore the spatial asymptotics of Eqs.~\eqref{eq:green_finite} and~\eqref{eq:green_finite_terms}. In the small $s$ limit, we find:
	\begin{equation}
		\displaystyle \hat{w}_\tau(s,q)\, \underset{s\rightarrow 0}{\sim}\, \frac{F_0 \Lambda}{4\pi G q}\sqrt{\frac{\mathcal{D}_\mathrm{pe}}{q}}\text{tanh}\left(\sqrt{\frac{q\tau^2}{\mathcal{D}_\mathrm{pe}}}\right).
	\end{equation}
	We note that the deformation $\hat{w}_\tau$ behaves as a constant with respect to the spatial frequency $s$ in the small-$s$ limit. However, the value of that constant depends on the temporal frequency $q$. Then, we compute the asymptotics in the large-$s$ limit, and find: 
	\begin{equation}
		\displaystyle \hat{w}_\tau(s,q)\, \underset{s\rightarrow \infty}{\sim}\, \frac{F_0 }{2\pi G s q}(1-\nu).
	\end{equation}
	In this limit, the deformation $\hat{w}_\tau$ decreases in a power law with a $-1$ exponent with respect to the spatial frequency $s$. Additionally, we recover the asymptotic result known for a purely elastic and compressible medium (see SI Sec.~\ref{sec:SI-asymptotic-elastic}). The two asymptotic behaviours in both large-$s$ and small-$s$ limits are indeed observed on Fig.~\ref{fig:green_reciprocal_time}A and B (inset). 
	\newline
	
	Furthermore, we observe a smooth transition between large-$s$ and small-$s$ regimes, in which the deformation reaches a maximum in reciprocal space (see Fig.~\ref{fig:green_reciprocal_time}). This observation is enhanced for larger temporal frequencies $q$ and happens around $s\sqrt{F_0/G}\sim 1$. Indeed, we refer to the asymptotics of both the incompressible and compressible responses from Eqs.~\eqref{asymptotereciprocal_finite} (see SI Sec.~\ref{sec:SI-asymptotic-elastic}) to characterize the transition. With $s$ increasing, we understand that the deformation behaves as a constant in the small-$s$ regime, then increases as a power law with exponent $2$ before reaching a maximum. Then, the deformation decreases according to a power law with a $-1$ exponent, switching from the incompressible to the compressible asymptote in the large-$s$ regime. Using tabulated inverse-Hankel transforms, one can build expectations on the asymptotic behaviour of the deformation profile in real space. In particular, the inverse Hankel transform of a constant is $\delta(r)/r$, the one of a square function $s^2$ is identically 0 and the one of an inverse function $1/s$ is $1/r$. Given these standard transforms, one can already predict that the deformation decreases as $1/r$ at small radii $r$, as for the purely elastic case, switching from a compressible to an incompressible behaviour with increasing $r$. With time, the deformation switches from an incompressible to a compressible behaviour, which means that the compressible behaviour is reached first at small radii and then propagates in time towards larger $r$. At large radii, the deformation vanishes completely. The transition between finite and zero deformations happens at about $r \sim \tau $, which means that the response of the material is suppressed beyond the peripheral region owing to the finite-size of the gel layer.
	\newline

	In the present subsection, we characterized the observed power-law behaviours and the transitions between limiting regimes at a fixed thickness, having chosen $\tau \sqrt{G/f_0} = 1$. For various temporal frequencies $q$, as for compressible and incompressible responses, the smooth transitions between the asymptotic behaviours at small and large $s$ are observed at a spacial frequency  of about $s\sqrt{F_0/G} \sim 1$. In the following subsection, we investigate the dependency on the thickness $\tau$. 
	
	\begin{figure}
		\centering
		\includegraphics[width=0.95\linewidth]{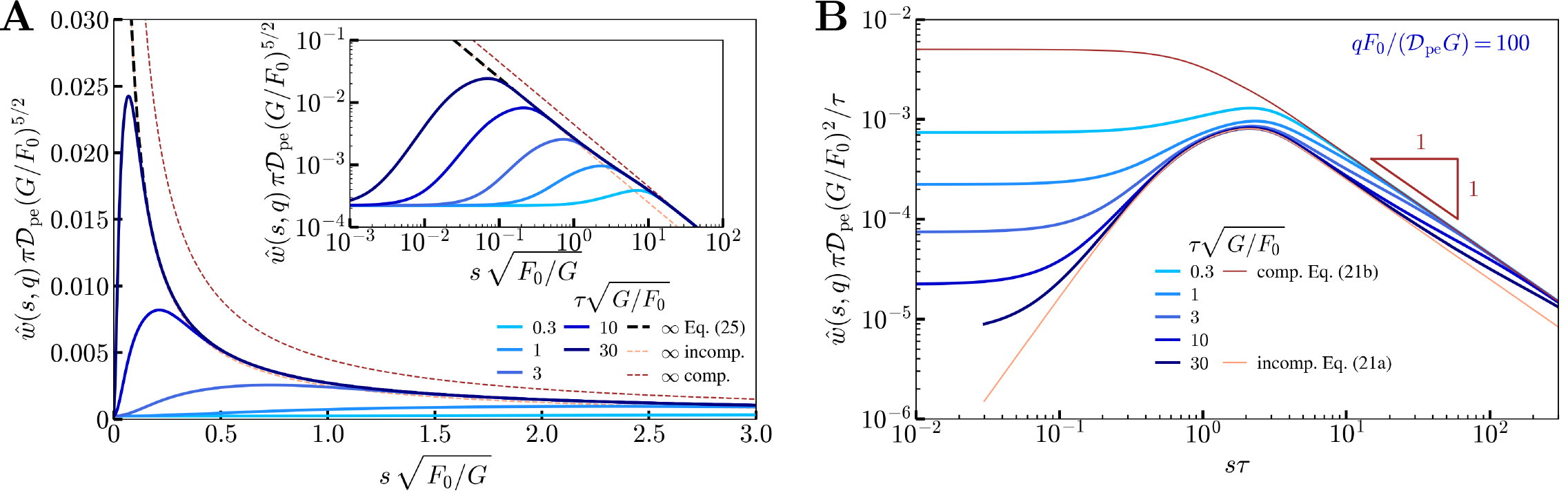}
		
		\caption{\textbf{Deformation profile of a finite-size poroelastic medium, for various thicknesses of the layer.} \textit{\textbf{A}: Deformation $\hat{w}_\tau (s,t)$ scaled by $(F_0/G)^{5/2}/(\pi\mathcal{D}_\text{pe})$, as a function of the dimensionless spatial frequency $s\sqrt{F_0/G}$, for different dimensionless thicknesses $\tau\sqrt{G/F_0}$, in linear (main panel) and logarithmic scales (inset). \textbf{B}: Ratio of the deformation to the thickness $\hat{w}_\tau/\tau$, scaled by  $(F_0/G)^{2}/(\pi\mathcal{D}_\text{pe})$, as a function of the rescaled variable $s\tau$, for various thicknesses dimensionless thicknesses $\tau\sqrt{G/F_0}$. The poroelastic, infinite-thickness response as defined in Eq.~\eqref{eq:def_poroelastic_infinite} is plotted in black dashed line~\cite{kopecz2023mechanical}. The known incompressible and compressible responses are plotted in orange and brown, respectively, for the infinite and finite-thickness cases, in dashed and full lines, respectively. The dimensionless temporal frequency is set at $qF_0/(\mathcal{D}_\text{pe}G) = 100$ and the Poisson ratio at $\nu = 0.1$. }}
		\label{fig:green_reciprocal_thickness}
	\end{figure}
	
	\subsection{From thick to thin films}
	\label{sec:thin_thick}
	We now investigate the effect of finite thickness on the poroelastic response. In Fig.~\ref{fig:green_reciprocal_thickness}A, we show the normalized deformation profiles for different thicknesses $\tau$. We observe that the response follows the reference infinite-thickness case in the large-$s$ regime, but deviates from the former reference in the small-$s$ regime. Additionally, the thicker the film, the lower the value of $s$ at which the transition occurs. To rationalize this observation, we explore the limit of infinite thickness, \textit{i.e.} when $\tau \rightarrow +\infty$. Doing so, we indeed recover the poroelastic response derived for an infinite medium in our previous work, ensuring the coherence of our calculations, as : 
	\begin{equation}
		\label{eq:def_poroelastic_infinite}
		\displaystyle  \lim_{\tau \rightarrow \infty\,}\, \hat{w}_\tau(s,q) =  \frac{F_0}{4\pi Gsq}\, \frac{1}{1 + \Lambda\chi\left(1-\beta\right)}  = 	\displaystyle \hat{w}_\infty(s,q).
	\end{equation}
	Moreover, we observe that the transition from the small-$s$ to the large-$s$ regimes is wider with increasing thickness, although still centred around $s\sqrt{F_0/G} \sim 1$. Thus, we chose to exhibit the deformation scaled by the thickness $\hat{w}_\tau / \tau$ as a function of the rescaled variable $\zeta = s \tau$ in Fig.~\ref{fig:green_reciprocal_thickness}B. In this representation, we observe again a transition region from the small-$s$ to the large-$s$ regimes in which the deformation reaches a maximum in reciprocal space, as in Fig.~\ref{fig:green_reciprocal_time}. This effect is enhanced for larger thicknesses. As a consequence, our prediction of the deformation profile behaviour in real space still holds. Additionally, we remark that the switch from the incompressible to the compressible asymptote is better observed for larger thicknesses at a given inverse time $q$. Not surprisingly, this means that for thicker gels, the relaxation to a compressible behaviour takes longer. To validate our predictions, we exhibit the deformation profile in real space in the next subsection.

	\subsection{Real-space deformation}
	\label{sec:real}
	\begin{figure}
	\centering
	\includegraphics[width=0.95\linewidth]{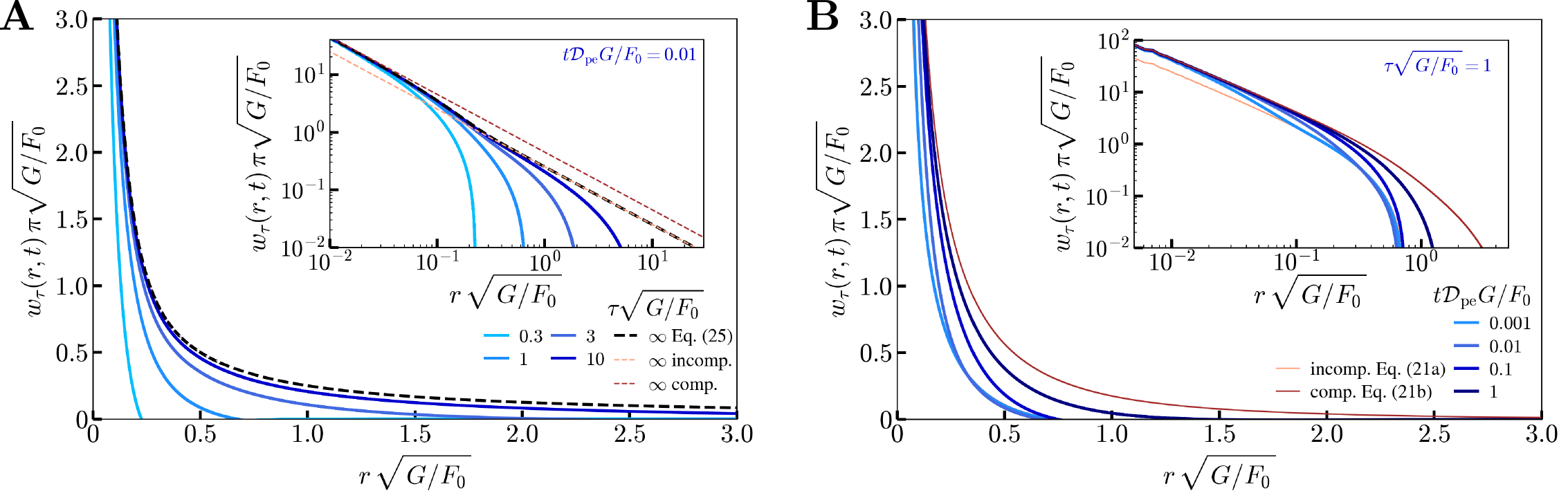}
	
	\caption{\textbf{Deformation profile of a poroelastic medium in real space. }\textit{\textbf{A}: Deformation $w_\tau (s,t)$ scaled by $\sqrt{F_0/G}/\pi$, as a function of the dimensionless radius $r\sqrt{G/F_0}$, for different dimensionless thicknesses $\tau\sqrt{G/F_0}$, in linear (main panel) and logarithmic scales (inset). The dimensionless time is set at $t \mathcal{D}_\text{pe}G/F_0 = 0.01$. The infinite-thickness response is represented with the black dashed line. The known incompressible and compressible responses for infinite-thickness elastic layers are plotted in orange and brown dashed lines, respectively. \textbf{B}: Deformation $w_\tau (s,t)$ scaled by $\sqrt{F_0/G}/\pi$, as a function of the dimensionless radius $r\sqrt{G/F_0}$, for different dimensionless times $t \mathcal{D}_\text{pe}G/F_0$, in linear (main panel) and logarithmic (inset) scales. The known incompressible and compressible responses for finite-thickness elastic layers are plotted in orange and brown full lines, respectively. The dimensionless thickness is set to $\tau\sqrt{G/F_0} = 1$ and the Poisson ratio to $\nu = 0.1$. }}
	\label{fig:green_real}
	\end{figure}
	
	To confirm our intuitive descriptions of the behaviour of the deformation profile in real space, we perform the inverse Laplace and Hankel transforms on the deformation profile, given by Eq.~\eqref{eq:green_finite_terms}. The result is not explicit, thus we use the Talbot algorithm to perform the inverse Laplace transform~\cite{abate2006unified}. To perform the inverse Hankel transform, reciprocal space is discretized on the zeros of Lagrange polynomials, such that integrals are defined using the Gauss-Legendre quadrature method and the transform is finally computed using Riemann summations~\cite{baddour2015theory}. We use 200 000 points in space both in Hankel and real spaces, on an finite reciprocal domain of $s\in[10^{-4}; 10^{2}]$ and using a uniform discretization of 0.05 step size in real space. Finally, to smoothen residual oscillations due to numerics, we apply a Savitsky-Golay filter of order 3 on a 9-point window in the data shown here.

	In panel A of Fig.~\ref{fig:green_real}, we show the deformation profile of a poroelastic layer resulting from a point force, at a given time for various thicknesses. In panel B, the deformation profile is plotted at various times for a given thickness. In contrast to the semi-infinite case, the main distinguishing feature of these finite-thickness cases is a decay to zero of the gel deformation at finite-radius. This zero crossing occurs at smaller radius for thinner films and for smaller times, with the cutoff being similar to the film thickness as predicted by the reciprocal-space asymptotics described in the previous section. For radii smaller than the cutoff, a transition between incompressible to compressible deformations occurs as described in detail next. 
	
	As also predicted from the explicit results in reciprocal space, we observe that the deformation decreases as $1/r$ at small radii $r$, which is reminiscent of the infinite-thickness case (Figs.~\ref{fig:green_real}A and B and Ref.~\cite{kopecz2023mechanical}). At large $r$, the decay switches from the compressible to the incompressible asymptote with increasing radius (Fig.~\ref{fig:green_real} A). Moreover, we observe a transition from the incompressible to the compressible asymptote with increasing time (Fig.~\ref{fig:green_real} B). In particular, at time $t=0$ the whole medium behaves incompressibly. Then, the transition to the compressible regime happens first in the region the closest to the excitation ---\textit{i.e} at small $r$--- then the transition progresses towards increasing radii with time. The latter transition is the manifestation of the poroelastic relaxation and constitutes the main feature of the poroelastic behaviour, as compared to purely elastic materials. In the opposite peripheral region, at larger radii $r$, we observe that the deformation completely vanishes as noted in the previous paragraph. The cut-off happens at a radius on the order of the thickness, and increases slightly with time up to a few times the thickness, until reaching the compressible asymptote.
	
	To summarize, the deformation of a finite-size poroelastic layer behaves similarly in space and time as the one of an infinite-thickness medium for radii $r\lesssim \tau$, and relaxes with a transition from an incompressible to a compressible purely elastic behaviour. On the contrary, for radii $r \gtrsim \tau$, the deformation vanishes, with a cut-off value that slightly increases in time, ranging from the known incompressible to the compressible purely elastic limits, which constitutes an important feature of finite-thickness poroelastic layers. 
	
	\subsection{Extension to axisymmetric pressure fields}
	\label{sec:extention}
		\begin{figure}[b!]
		\centering
		\includegraphics[width=0.75\linewidth]{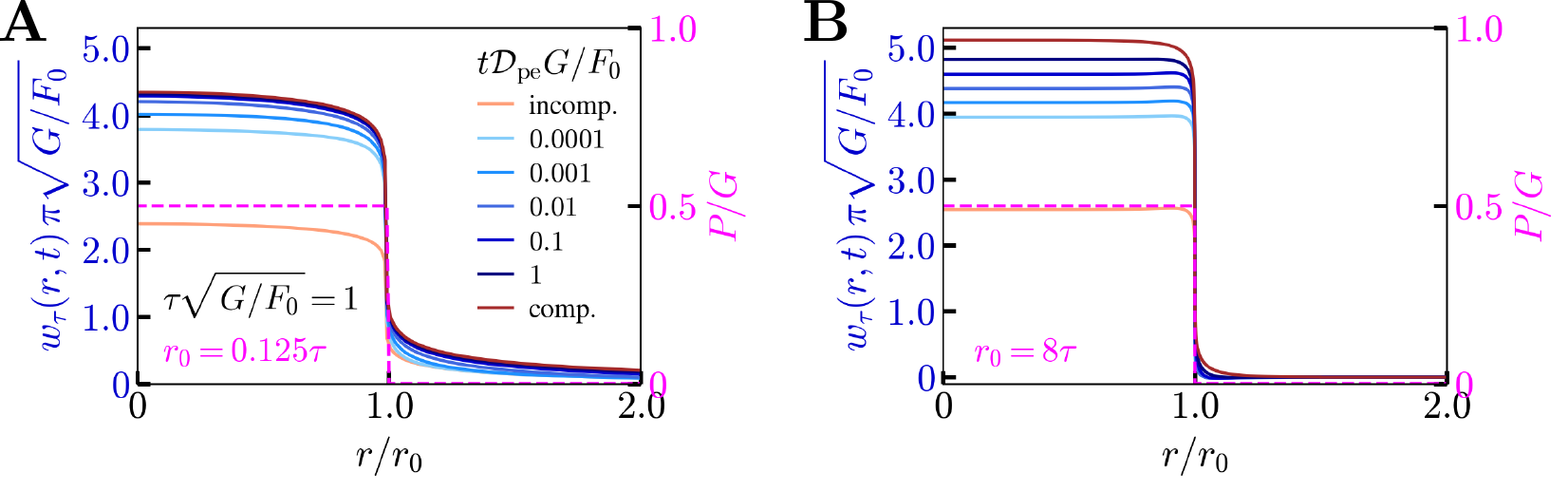}
		
		\caption{\textbf{Example: relaxation of the deformation profile generated by a gate-shaped pressure field. }\textit{Deformation $w_\tau(r,t)$ scaled by $\sqrt{F_0/G}/\pi $, as a function of the radius $r$ scaled by the extent radius of the pressure field $r_0$, for different dimensionless times $t \mathcal{D}_\text{pe}G/F_0$. \textbf{A}: The radial extent of the pressure field is fixed to $r_0 = \tau/8$. \textbf{B}: Same with $r_0 = 8\tau$. The orange and brown lines represent respectively the initial and final responses. The dimensionless thickness is set to $\tau\sqrt{G/F_0} =1 $ and the Poisson ratio to $\nu = 0.1$.}}
		\label{fig:hat}
	\end{figure}
	
Until the present subsection, we only studied the deformation profile resulting from a point-force pressure source. However, real experiments and realistic modelling involve finite-width force profiles, \emph{e.g.} from an indenter. Using the superposition theorem, and within the linear-response approximation, the results presented in this article can be generalized to any axisymmetric pressure field, using the following convolution product : 
	\begin{equation}
		\displaystyle w(\text{\textbf{r}},t) = \int_{-\infty}^t \,\text{d}t'\int_{\mathbb{R}} \,\text{d}^2\text{\textbf{r}}' \mathcal{G}_\tau (|\text{\textbf{r}} -\text{\textbf{r}}'|, t-t')P(\bm{r}',t'). 
	\end{equation}
	Here, we use the finite-size poroelastic Green's function of the problem as given in Eq.~\eqref{eq:green_finite_def}, and we use the inverse Hankel and Laplace transforms, as:
	\begin{equation}
		\displaystyle \mathcal{G}_\tau (r,t) = \frac{1}{2\pi i}  \int_{\gamma -i \infty}^{\gamma +i\infty} \text{d}q \, e^{q t} \int_0^\infty \text{d}s\, \hat{\mathcal{G}}_\tau(s,q)  s J_0(sr).
		\label{eq:green-convolution}
	\end{equation}

	The latter generalization to axisymmetric situations highlights the relevance of using the Green's function for modelling any approach or indentation experiments conducted on soft and porous matter in the same symmetry conditions. As a demonstration, we exhibit here a simple example; namely, the relaxation in time of the deformation profile induced by a gate-shaped pressure field is considered, given by the pressure profile $P(r)$ turned on at $t = 0^+$ specified by: 
	\begin{equation} 
		\displaystyle P(r) = \left\{
		\begin{array}{l}
		\displaystyle  p_0 \quad \text{for} \quad r\leq r_0\\
		\displaystyle 0  \quad \text{for} \quad r > r_0.\\
		\end{array}
		\right.
	\end{equation}
	The deformation profile in time is computed numerically using Eq.~\eqref{eq:green-convolution} and using the numerically-inverted Green's function shown in Fig.~\ref{fig:green_real}. The deformation as a function the the radius $r$ scaled by the radial extent of the pressure field $r_0$ is shown for different dimensionless times in Fig.~\ref{fig:hat}. Two cases are considered: (i) the radial extent of the pressure field is an eighth of the poroelastic film thickness (Fig.~\ref{fig:hat}A) and (ii) the radial extent of the pressure field is eight times the film thickness (Fig.~\ref{fig:hat}B). We observe that the deformation profile reaches its final shape faster for a narrower pressure field. Additionally, the deformation-profile shape resembles more the pressure-stimulus shape for a wider pressure field, with a smoothening of the pressure discontinuity for narrower inputs.
	\newline
	
	Besides the simple gate-like function, the above framework applies also to, and for example, colloidal-probe atomic force microscopy~\cite{butt1991measuring,ducker1991direct}, indentation tests~\cite{hu2010using, delavoipiere2016poroelastic}, experiments conducted in sphere-plane surface forces apparatus~\cite{israelachvili2010recent,degen2020poroelasticity,crassous1993experimental}, as well as experiments performed with home-made spherical probes~\cite{wang2015out, wang2017elastic2,wang2017elastic}.
	
	\section{Conclusion}
	
	We theoretically addressed the mechanical response of a permeable poroelastic layer of finite thickness, submitted to a point-force pressure source, in the linear regime of small deformations. We first computed the fundamental point-force response and analytically derived the Green's function of the problem. As in our previous work on semi-infinite media, we showed that the main feature of poroelasticity is a relaxation from incompressible to compressible, and purely elastic behaviours. This relaxation progresses in time and space diffusively. However, in contrast to semi-infinite media, here we showed that the main feature of finite thicknesses is that the mechanical response of the layer extends around the application point only to a radius on the order of the poroelastic layer thickness. The latter thickness defines then the appropriate scale to study the relaxation of poroelastic layers. Finally, by convolution, the fundamental response expressed by the Green's function is extended to any axisymmetric problem. As an illustration, we exhibited the simple example of a gate-shaped pressure field. Our work may find applications in the investigation of the mechanical properties of soft and thin gels such as polymeric coatings and biological tissues or membranes, and sets the relevant spatial scale and theoretical framework for such studies. 
	\newline
	
	\noindent \textbf{Data availability:} The data shown in the present article are available from the authors upon reasonable request.\\
	\noindent \textbf{Conflict of interest:} The authors have no conflit of interest to declare.\\
	\noindent \textbf{Funding:} The authors acknowledge financial support from the European Union through the European Research Council under EMetBrown (ERC-CoG-101039103) grant. Views and opinions expressed are however those of the authors only and do not necessarily reflect those of the European Union or the European	Research Council. Neither the European Union nor the granting authority can be held responsible for them. The authors also acknowledge financial support from the Agence Nationale de la Recherche under EMetBrown (ANR-21-ERCC-0010-01), Softer (ANR21-CE06-0029), CoPinS (ANR-19-CE06-0021), and Fricolas (ANR-21-CE06-0039) grants, as well as from the Interdisciplinary and Exploratory Research program at Univ. Bordeaux under MISTIC grant, France. Besides, they acknowledge the support from the Réseau de Recherche Impulsion (RRI) “Frontiers of Life”, which received financial support from the French government in the framework of Univ. Bordeaux’s France 2030 program. Finally, they thank the Soft Matter Collaborative Research Unit, Frontier Research Center for Advanced Material and Life Science, Faculty of Advanced Life Science at Hokkaido University, Sapporo, Japan, and the CNRS International Research Network between France and India on “Hydrodynamics at small	scales: from soft matter to bioengineering”.\\
	\noindent \textbf{Contributions:} CKM: formal analysis, investigation, visualization, writing, review and editing; JDM: conceptualization, funding acquisition, methodology, project administration, resources, supervision, writing—review and editing; TS: conceptualization, funding acquisition, methodology, project administration, resources, supervision, writing—review and editing.\\
	\noindent \textbf{Acknowledgments:} The authors gratefully thank Vincent Bertin and Elisabeth Charlaix for fruitful discussions, and Daniel Acuña for access to the calculation cluster. 
	
	\bibliographystyle{ieeetr}
	\bibliography{Kopecz-Muller2026.bib}

	\newpage
	
	\section*{Supplementary Information}
	\setcounter{equation}{0}
	\setcounter{section}{0}
	\setcounter{subsection}{0}
	\renewcommand{\theequation}{S\arabic{equation}}
	\renewcommand{\thesubsection}{SI-\arabic{subsection}}

	\subsection{Detailed resolution}
	\label{sec:SI-detail-calculation}
	In this section, we detail the calculation path to compute the finite-size poroelastic Green's function, which is briefly summarized in Sec.~\ref{sec:resolution} and reported in~\cite{kopecz2024mechanics}.
	
	A first step consists in reducing the problem to a system of two coupled equations on the solvent concentration and the chemical potential. We first combine Eqs.~\eqref{eq:stress-strain} and~\eqref{eq:navier} using Eqs.~\eqref{eq:strain-def} and~\eqref{eq:div-def}, which leads to:
	\begin{equation}
		\displaystyle G\Omega\left(\bm{\nabla}^2 \mathbf{u} + \frac{\Omega}{1-2\nu} \bm{\nabla} \left(c-c_0\right)\right) = \bm{\nabla}\left(\mu - \mu_0\right).
		\label{eq:u-c-mu}
	\end{equation}
	Taking the gradient of Eq.~\eqref{eq:u-c-mu} and using again Eq.~\eqref{eq:div-def}, we obtain a first equation on the chemical potential and solvent concentration, that reads:
	\begin{equation}
		\displaystyle 2G\Omega^2 \frac{1-\nu}{1-2\nu}\bm{\nabla}^2\left(c-c_0 \right) = \bm{\nabla}^2\left(\mu-\mu_0\right).
		\label{eq:c-mu}
	\end{equation}
	Then, we combine Eqs.~\eqref{eq:darcy} and~\eqref{eq:adv} to obtain a second equation on the chemical potential and solvent concentration:
	\begin{equation}
		\displaystyle \frac{\partial c }{\partial t} = \left(\frac{k}{\eta \Omega^2}\right) \bm{\nabla}^2 \mu.
		\label{eq:c-mu2}
	\end{equation}
	Using Eq.~\eqref{eq:c-mu}, we finally obtain a system of two equations on the chemical potential $\mu$ and the solvent concentration $c$, which reads:
		\begin{subequations}
		\begin{align}
			\label{eq:system-c-mu1}
			\displaystyle \bm{\nabla}^2 \left[\left(\mu-\mu_0\right) - 2G\Omega^2 \frac{1-\nu}{1-2\nu}\left(c-c_0\right)\right]=0\\
			\label{eq:system-c-mu2}
			\displaystyle \mathcal{D}_\text{pe}  \bm{\nabla}^2c -  \frac{\partial c }{\partial t} =0.
		\end{align}
		\label{eq:system-c-mu}
	\end{subequations}
	We define the poroelastic effective diffusion coefficient as:
	\begin{equation}
		\displaystyle 	\mathcal{D}_\text{pe}  = \frac{2}{\Lambda} \frac{Gk}{\eta},
	\end{equation}
	recalling that the compressibility factor is defined in Eq.~\eqref{eq:LambdaDef} as $\Lambda = (1-2\nu)/(1-\nu)$.

	A second step aims to decouple the system of equations defined in Eq.~\eqref{eq:system-c-mu}. We define two displacement potential functions $A(r,z,t)$ and $B(r,z,t)$, as~\cite{mcnamee1960displacement,mcnamee1960plane}: 
	\begin{subequations}
		\begin{align}
			\displaystyle u_r &= z\frac{\partial A}{\partial r}+ \frac{\partial B}{\partial r}, \\
			\displaystyle u_z &=	z\frac{\partial A}{\partial z}-A + \frac{\partial B}{\partial z}.
		\end{align}
		\label{eq:disp-potential}
	\end{subequations}
	The two potential function should satisfy the following equations:
		\begin{subequations}
		\begin{align}
			\label{eq:mcnamee1}
			\displaystyle 2G\Omega\frac{\partial A}{\partial z} &= \left(\mu - \mu_0\right) -  \frac{2G\Omega^2}{\Lambda}\left(c-c_0\right),\\
			\label{eq:mcnamee2}
			\displaystyle \bm{\nabla}^2 B &= \Omega\left(c-c_0\right),\\
			\label{eq:mcnamee3}
			\displaystyle \bm{\nabla}^2 A &= 0,\\
			\label{eq:mcnamee4}
			\displaystyle \mathcal{D}_\text{pe} \bm{\nabla}^4B &= \frac{\partial \bm{\nabla}^2B}{\partial t}.
		\end{align}
	\end{subequations}
	One can note that the first two lines respectively define the quantities to replace in Eqs.~\eqref{eq:system-c-mu1} and~\eqref{eq:system-c-mu2}, while the last two lines respectively result from applying the replacement. Additionally, the chemical potential and the $z$-components of the stress tensor can be expressed as functions of the potentials, as follows:
	\begin{subequations}
		\begin{align}
			\mu-\mu_0 &= 2G\Omega \left[\frac{\partial A}{\partial z} + \frac{1}{\Lambda}\bm{\nabla}^2B\right],\\
			\sigma_{zz} &= 2G\left[z\frac{\partial^2 A}{\partial z^2} - \frac{\partial A}{\partial z}  + \frac{\partial^2 B}{\partial z^2} -\bm{\nabla}^2B  \right],\\
			\sigma_{rz} &= 2G\left[z\frac{\partial^2 A}{\partial r \partial z} +   \frac{\partial^2 B}{\partial z \partial r} \right].
		\end{align}
		\label{eq:potential-sigma-AB}
	\end{subequations}
	
	In a third step, we switch to reciprocal space, using the Hankel and Laplace transforms introduced in Eq.~\eqref{eq:space_forward}. The potential function $A$ and $B$ are transformed using the zeroth-order Hankel transform. Equations~\eqref{eq:mcnamee3} and~\eqref{eq:mcnamee4} lead to the following linear ordinary differential equations:
	\begin{subequations}
		\begin{align}
			\displaystyle \left(\frac{\partial ^2 }{\partial z^2} - s^2 \right) \hat{A} &= 0, \\
			\displaystyle \left(\frac{\partial ^2 }{\partial z^2} - s^2 - \frac{q}{\mathcal{D}_\text{pe}} \right)\left(\frac{\partial ^2 }{\partial z^2} - s^2 \right) \hat{B} &= 0.
		\end{align}
		\label{eq:linear-ordinary}
	\end{subequations}
	In the present case of a finite-thickness gel, the general solutions of Eqs.~\eqref{eq:linear-ordinary} are:
	\begin{subequations}
		\begin{align}
			\displaystyle \hat{A} &= a_1e^{sz} + a_2e^{-sz},\\
			\displaystyle \hat{B} &= b_1e^{sz} +b_2e^{sz\beta}+ b_3e^{-sz}+b_2e^{-sz\beta},
		\end{align}
		\label{eq:sol-general}
	\end{subequations}
where $\beta$ is defined in Eq.~\eqref{eq:dimensionless_variable_finite_beta}. The six integration constants $a_1$, $a_2$, $b_1$, $b_2$, $b_3$, $b_4$ that depend on the spatial and temporal frequencies $s$ and $q$ are determined using the boundary conditions, as given in Sec.~\ref{sec:BC}. To perform this calculation, the displacement field components, chemical potential and stress tensor components must be expressed in reciprocal space as functions of the transformed potential functions $\hat{A}(s,z,q)$ and $\hat{B}(s,z,q)$. Using their expressions as in Eqs.~\eqref{eq:disp-potential} and~\eqref{eq:potential-sigma-AB} and transforming them using Eq.~\eqref{eq:space_forward}, one gets:
	\begin{subequations}
		\begin{align}
			\displaystyle \hat{u_s}(s,z,q) &= -zs\hat{A} -s\hat{B}\\
			\displaystyle \hat{u_z}(s,z,q) &= z \frac{\partial \hat{A}}{\partial z} - \hat{A} + \frac{\partial \hat{B}}{\partial z}\\
			\displaystyle  \hat{\mu}(s,z,q)	&= 2G\Omega\left[\frac{\partial \hat{A}}{\partial z} + \frac{1}{\Lambda}\left(\frac{\partial ^2 }{\partial z^2} - s^2 \right) \hat{B} \right]\\
			\displaystyle \hat{\sigma}_{zz}(s,z,q) &= 2G \left[z\frac{\partial^2 \hat{A}}{\partial z^2} - \frac{\partial \hat{A}}{\partial z}  + s^2 \hat{B}\right]\\
			\displaystyle  \hat{\sigma}_{sz}(s,z,q) &= -2Gs \left[z\frac{\partial \hat{A}}{\partial z} +\frac{\partial \hat{B}}{\partial z}\right].
		\end{align}
	\end{subequations}
	Expressing the boundary conditions both at the free interface ($z=0$) and at the bottom of the gel ($z=-\tau$), and using Eqs.~\eqref{eq:sol-general}, we obtain a set of six equations on the six integration contants, as:
	\begin{subequations}
		\label{eq:BC_transformed_finite}
		\begin{align}
			\displaystyle \hat{\sigma}_{sz}(s,z=0,q) &= \quad 0\quad = -2Gs^2\left[b_1 -b_3 + (b_2-b_4)\beta\right],  \\
			\displaystyle\hat{\sigma}_{zz}(s,z=0,q) &=-\frac{ F_0}{2\pi q} = 2Gs^2\left[-\frac{a_1}{s}+\frac{a_2}{s} +(b_1+b_2+b_3+b_4)\right],\\
			\displaystyle \hat{\mu}(s,z=0,q)- \hat{\mu}_0&= \quad 0\quad=  2G\Omega s^2\left[\frac{a_1}{s} - \frac{a_2}{s} + \frac{1}{\Lambda \chi}(b_2+b_4)\right],\\
			\displaystyle \hat{u}_s(s,z=-\tau,q) &=\quad 0\quad =  -s\left[-\tau\Big(a_1e^{-\zeta}+a_2e^{\zeta} \Big) + b_1e^{-\zeta}+b_2e^{-\zeta\beta} + b_3e^{\zeta}+b_4e^{\zeta\beta}\right],\\
			\displaystyle \hat{u}_z(s,z=-\tau,q)& = \quad 0\quad= -s\left[\frac{1+\zeta}{s}a_1e^{-\zeta} +\frac{1-\zeta}{s}a_2e^{\zeta} -e^{-\zeta}b_1 - \beta b_2 e^{-\zeta\beta} + e^{\zeta}b_3 +\beta b_4 e^{\zeta\beta}  \right],\\
			\displaystyle \frac{\partial \mu}{\partial z}(s,z=-\tau,q) &=\quad 0\quad= 2G\Omega s^3\left[\frac{a_1}{s}e^{-\zeta}+ \frac{a_2}{s}e^{\zeta} + \frac{\beta}{\Lambda \chi}b_2 e^{-\zeta\beta} -  \frac{\beta}{\Lambda \chi}b_4 e^{\zeta\beta} \right], \label{eq:def_zeta}
		\end{align}
	\end{subequations}
	using the auxiliary variables defined in~\eqref{eq:dimensionless_variable_finite}. Finally, Eq.~\eqref{eq:BC_transformed_finite} is solved using a formal calculation software and the potential functions $\hat{A}$ and $\hat{B}$ are fully determined. 
	\newline
	
	\subsection{Non-dimensionalization}
	\label{sec:SI-non-dim}
	In this section, we detail the switch to dimensionless variables and justify the scalings that appear in Figs.~\ref{fig:green_reciprocal_time},~\ref{fig:green_reciprocal_thickness} and~\ref{fig:green_real}. For this purpose, let us write a tilde above dimensionless quantities.
	
	First, the radius is scaled by $\sqrt{F_0/G}$ and the time by $F_0/(G \mathcal{D}_\text{pe})$. In reciprocal space, the spatial and temporal frequencies scale with the inverses of their homologues in real space. One gets:
	\begin{equation}
		\displaystyle \tilde{r} = r\sqrt{\frac{G}{F_0}}, \quad \tilde{t} = t \frac{G \mathcal{D}_\text{pe}}{F_0},  \quad \tilde{s} =  s\sqrt{\frac{F_0}{G}},  \quad  \text{and}  \quad \tilde{q} = q\frac{F_0}{G \mathcal{D}_\text{pe}}.
	\end{equation}
	In real space, the deformation is a length. The dimensionless deformation reads : 
	\begin{equation}
		\displaystyle \tilde{w} = w \sqrt{\frac{G}{F_0}}.
	\end{equation}
	However, the Laplace and Hankel transforms modify the dimension of the transformed object. For the deformation, one has :
	\begin{align}
		\begin{split}
			\displaystyle \hat{w}(s,q) &=  \int_{0}^{\infty} \text{d}t \, e^{-q t}\int_0^\infty \text{d}r\text{ } w(r,t)  r J_0(sr) \\
			\displaystyle 	&=  \int_{0}^{\infty}   \text{d}\tilde{t}\, \frac{F_0}{G \mathcal{D}_\text{pe}} e^{-\tilde{q}\tilde{t}} \int_0^\infty \text{d}\tilde{r} \sqrt{\frac{F_0}{G}}^3 \tilde{w}(\tilde{r},\tilde{t}) \tilde{r}J_0(\tilde{s}\tilde{r})\\
			\displaystyle 	&= \frac{1}{\mathcal{D}_\text{pe}}\left(\frac{F_0}{G}\right)^{5/2}\int_{0}^{\infty}   \text{d}\tilde{t}\,e^{-\tilde{q}\tilde{t}}\int_0^\infty \text{d}\tilde{r}  \tilde{w}(\tilde{r},\tilde{t}) \tilde{r}J_0(\tilde{s}\tilde{r})\\
			& =  \frac{1}{\mathcal{D}_\text{pe}}\left(\frac{F_0}{G}\right)^{5/2} \tilde{\hat{w}}(\tilde{s},\tilde{q}).
		\end{split}
	\end{align}
	As a consequence, the deformation is scaled by $\left(F_0/G\right)^{5/2}/\mathcal{D}_\text{pe} $ in reciprocal space. In Figs.~\ref{fig:green_reciprocal_time} and~\ref{fig:green_reciprocal_thickness}, it is represented as scaled by  $\left(F_0/G\right)^{5/2}/(\pi\mathcal{D}_\text{pe}) $ for convenience. 
	\newline
	
		\subsection{Central and peripheral asymptotics of the purely elastic case}
	\label{sec:SI-asymptotic-elastic}
	To understand the behaviour of the deformation profiles as a transition from the incompressible to the compressible asymptotes, we now compute the asymptotics of the latter, given by Eqs.~\eqref{asymptotereciprocal_finite} in the small- and large-$s$ limits, as:
	\begin{subequations}
		\label{eq:asymtotics_finite}
		\begin{align}
			\label{eq:asymtotics_imcomp_smallS}
			\displaystyle \hat{w}_\tau^{\mathrm{incomp}}(s)\, \underset{s\rightarrow 0}{\sim}\,&\, \frac{F_0 \tau^3}{6\pi G}s^2,\\
			\label{eq:asymtotics_imcomp_largeS}
			\displaystyle \hat{w}_\tau^{\mathrm{incomp}}(s) \underset{s\rightarrow \infty}{\sim}&\, \frac{F_0}{4\pi Gs},\\
			\label{eq:asymtotics_comp_smallS}
			\displaystyle \hat{w}_\tau^{\mathrm{comp}}(s)\, \underset{s\rightarrow 0}{\sim}\,&\, \frac{F_0 \tau}{4\pi G}\,\frac{1-2\nu}{1-\nu},\\
			\label{eq:asymtotics_comp_largeS}
			\displaystyle \hat{w}_\tau^{\mathrm{comp}}(s) \underset{s\rightarrow \infty}{\sim}&\, \frac{F_0(1-\nu)}{2\pi Gs}.
		\end{align}
	\end{subequations}

	Using the inverse Laplace and Hankel transforms as defined in Eq.~\eqref{eq:inversion} on the asymptotics written in Eqs.~\eqref{eq:asymtotics_finite}, we get an estimate of the purely-elastic response behaviour in real space, as:
	\begin{subequations}
		\label{eq:asymtotics_finite_real}
		\begin{align}
			\label{eq:asymtotics_imcomp_largeR}
			\displaystyle w_\tau(r,t=0^+) = w_\tau^{\mathrm{incomp}}(r) \underset{r\rightarrow \infty}{\sim}&\, 0,\\
			\label{eq:asymtotics_imcomp_smallR}
			\displaystyle w_\tau(r,t=0^+) =  w_\tau^{\mathrm{incomp}}(r)\, \underset{r\rightarrow 0}{\sim}\,&\, \frac{F_0}{4\pi Gr},\\
			\label{eq:asymtotics_comp_largeR}
			\displaystyle w_\tau(r,t\rightarrow\infty) = \,\, w_\tau^{\mathrm{comp}}(r)\,\, \underset{r\rightarrow \infty}{\sim}&\, \frac{F_0 \tau}{4\pi G}\,\frac{1-2\nu}{1-2\nu} \frac{\delta(r)}{r} \underset{r\rightarrow \infty}{\sim}\,0,\\
			\label{eq:asymtotics_comp_smallR}
			\displaystyle  w_\tau(r,t\rightarrow\infty) =\,\, w_\tau^{\mathrm{comp}}(r)\,\,\, \underset{r\rightarrow 0}{\sim}\,&\, \frac{F_0(1-\nu)}{2\pi Gr},
		\end{align}
	\end{subequations}
	recalling that the inverse Hankel transforms of a constant is $\delta(r)/r$ and the one of the square function $s^2$ is identically 0.
	\newline

\end{document}